\documentclass[prd,showpacs,preprintnumbers,amsmath,amssymb,superscriptaddress,floatfix,nofootinbib]{revtex4}
\usepackage{graphicx}                      
\usepackage{epstopdf}
\usepackage{amsmath} 
\usepackage{amsfonts}
\usepackage{amssymb}
\usepackage{bm}

\begin{document}

\title{Theoretical study of the $D^0 \to K^- \pi^+ \eta$ reaction.}
\author{Genaro Toledo }
\email{toledo@fisica.unam.mx}
\affiliation{Instituto de F\'{\i}sica, Universidad Nacional Aut\'onoma de M\'exico, AP 20-364, Ciudad de M\'exico 01000, M\'exico.}

\affiliation{Departamento de F\'{\i}sica Te\'orica and IFIC,
Centro Mixto Universidad de Valencia-CSIC Institutos de Investigaci\'on de Paterna, Aptdo.22085, 46071 Valencia, Spain}

\author{Natsumi Ikeno}
\email{ikeno@tottori-u.ac.jp}
\affiliation{
Department of Agricultural, Life and Environmental Sciences, Tottori University, Tottori 680-8551, Japan}
\affiliation{Departamento de F\'{\i}sica Te\'orica and IFIC,
Centro Mixto Universidad de Valencia-CSIC Institutos de Investigaci\'on de Paterna, Aptdo.22085, 46071 Valencia, Spain}

\author{Eulogio Oset}
\email{oset@ific.uv.es}
\affiliation{Departamento de F\'{\i}sica Te\'orica and IFIC,
Centro Mixto Universidad de Valencia-CSIC Institutos de Investigaci\'on de Paterna, Aptdo.22085, 46071 Valencia, Spain}
\date{\today} 

\begin{abstract}
\noindent	
  We develop a model to study the $D^0 \to K^- \pi^+ \eta$ weak decay, starting with the color favored external emission and Cabibbo favored mode at the quark level. A less favored internal emission decay mode is also studied as a source of small corrections. Some pairs of quarks are allowed to hadronize producing two pseudoscalar mesons, which posteriorly are allowed to interact to finally provide the $K^- \pi^+ \eta$ state. The chiral unitary approach is used to take into account the final state interaction of pairs of mesons, which has as a consequence the production of the $\kappa$  ($K^*_0(700)$) and the $a_0(980)$ resonances, very well visible in the invariant mass distributions. We also introduce the $\bar{K}^{*0} \eta$ production in a phenomenological way and show that the $s$-wave pseudoscalar interaction together with this vector excitation mode are sufficient to provide a fair reproduction of the experimental data. The agreement with the data, in particular the relative weight of the $a_0(980)$ to the $\kappa$ excitation, provides extra support to the picture used, in which these two resonances are a consequence of the interaction of pseudoscalar mesons and not ordinary $q \bar{q}$ mesons. 

\end{abstract}



\maketitle

\section{Introduction}
The weak decay of heavy mesons into several mesons has received much attention in the past and continues to draw attention nowadays. In particular, three meson decays of $D$ mesons already captured attention in early days, looking at the topology of the decay at the quark level and the posterior hadronization of  pairs of quarks into mesons \cite{ellis,matsuda,nakagawa}.  More recently the emphasis is put in the valuable information that these processes contain on the final state interaction of pairs of mesons and the production of resonances \cite{review}. The existence of three particles in the final state gives much flexibility to play with the invariant mass of pairs of particles, providing ranges where several resonances appear. The Dalitz plot and the projected invariant mass distributions are thus very rich, containing much information on the dynamics of mesons.
   In this direction, the data on the $D^+ \to \pi^+ \pi^- \pi^+$  reaction  are used in \cite{aitala} to determine parameters for the $f_0(980)$ and $f_0(1370)$. Further steps in this direction analyzing the invariant mass spectra in the $D^+$ and $D_s^+$ decay into three pions are given in \cite{focus} using the $K$-matrix approach to deal with the $\pi-\pi$ interaction. The same Dalitz plot distributions are analyzed in \cite{klempt} using different partial wave analysis within the $K$-matrix approach, trying to extract information on different scalar meson states. An interesting feature appears in the $D^0 \to \pi^+ \pi^- \pi^0$ reaction measured by the BaBar collaboration \cite{babar,babar2} where the final pion pairs are surprisingly dominated by isospin $I=0$, but this feature, rather than being tied to a dynamical property of the final state interaction, was found to be a consequence of subtle cancellations between different topological decay modes entering the reaction \cite{rosner}. The $D^+ \to K^- \pi^+ \pi^+$ ($D^0 \to K^0_s \pi^+ \pi^+$) decay mode \cite{muramatsu,kappa}  was also instrumental in this direction, showing a clear signal for the $\kappa$ resonance ($K^*_0(700)$) in the $\pi K$ channel, which was analyzed in detail in \cite{ollerkappa} within the chiral unitary approach, and later on in \cite{patricia}. A different approach to that reaction is followed in \cite{kubis} by means of dispersion relations and input of experimental phase shifts as a way to take into account the final state interaction of the meson components. The related $D^0 \to K^0 \pi^+ \pi^-$ reaction was also the object of a detailed study considering the final state interaction by means of amplitudes tested in other reactions \cite{robert}. Similarly the $D^+ \to K^+ K^- K^+$ has been also thoroughly studied in \cite{dosreis} with the aim of obtaining information on the $K \bar{K}$ interaction.\\ 

   The advent of the chiral unitary approach for the meson meson interaction \cite{npa,ramonet,kaiser,markushin,juan} has brought new tools to analyze these reactions, allowing one to make predictions for mass distributions with a minimum input. The agreement found with the data serves in most cases to support the dynamical character of some resonances, which appear as a consequence of the meson meson interaction and are not of $q \bar{q}$ nature. In this line the $D^0$ decays to  $K^0_S$ plus $f_0(500)$, $f_0(980)$ or $a_0(980)$ were studied in \cite{daixie}, and the relative strength for the excitation of these resonances was predicted in that scheme, showing agreement with experiment in the ratios available.  In \cite{dias} the $D_s^+ \to \pi^+ \pi^- \pi^+$ and $\pi^+ K^+ K^-$ decays were studied and the role of the $f_0(980)$ resonance in the $\pi^+ \pi^-$ and $K^+ K^-$ mass distributions was established. In \cite{sakailiang} the $D_s^+ \to \pi^+ \pi^0$ plus $a_0(980)$ or $f_0(980)$ reactions were studied and, thanks to the presence of a triangle singularity, an abnormal isospin violation was found with large mixing of the two scalar resonances. One of the findings of the chiral unitary approach in the meson sector is the existence of two $K_1(1270)$ resonances \cite{luis,geng}, much in resemblance with the two $\Lambda(1405)$ states \cite{ollerulf,cola,juancarmen,hyodo,miyahara}. Taking this into account, predictions for the production of these two resonances were done in \cite{guangying} in the decay of $D^0 \to \pi^+$ plus $\rho K$ or $K^* \pi$. Finally, in \cite{molina} the $D_s^+ \to \pi^+ \pi^0 \eta$ reaction measured by the BESIII collaboration \cite{besiii} was studied and a good agreement with data was found, showing that the mechanism for production was internal emission rather than annihilation as suggested in the experimental paper.\\ 

    The reaction that we study here, the $D^0 \to K^- \pi^+ \eta$ decay, measured by the Belle collaboration \cite{belle}, is similar to the latter one mentioned above, but in addition to the $\pi \eta$ interaction which leads to the $a_0(980)$ resonance, here one also has the  $K \pi$ interaction, which shows as a $p$-wave resonance in the form of a $K^*$, and also in $s$-wave, giving rise to the $\kappa$ ($K^*_0(700)$), which are both well visible in the data. Our study, using the chiral unitary approach, shows how the two $s$-wave signals are related in the theoretical scheme and comparison with the data allows a theoretical interpretation of the results, showing the value of the reaction to provide information on the meson meson interaction and indirectly on the nature of the $a_0(980)$ and  $K^*_0(700)$ resonances.\\

\section{Formalism}
\begin{figure}[b!]
\begin{center}
\includegraphics[scale=0.4]{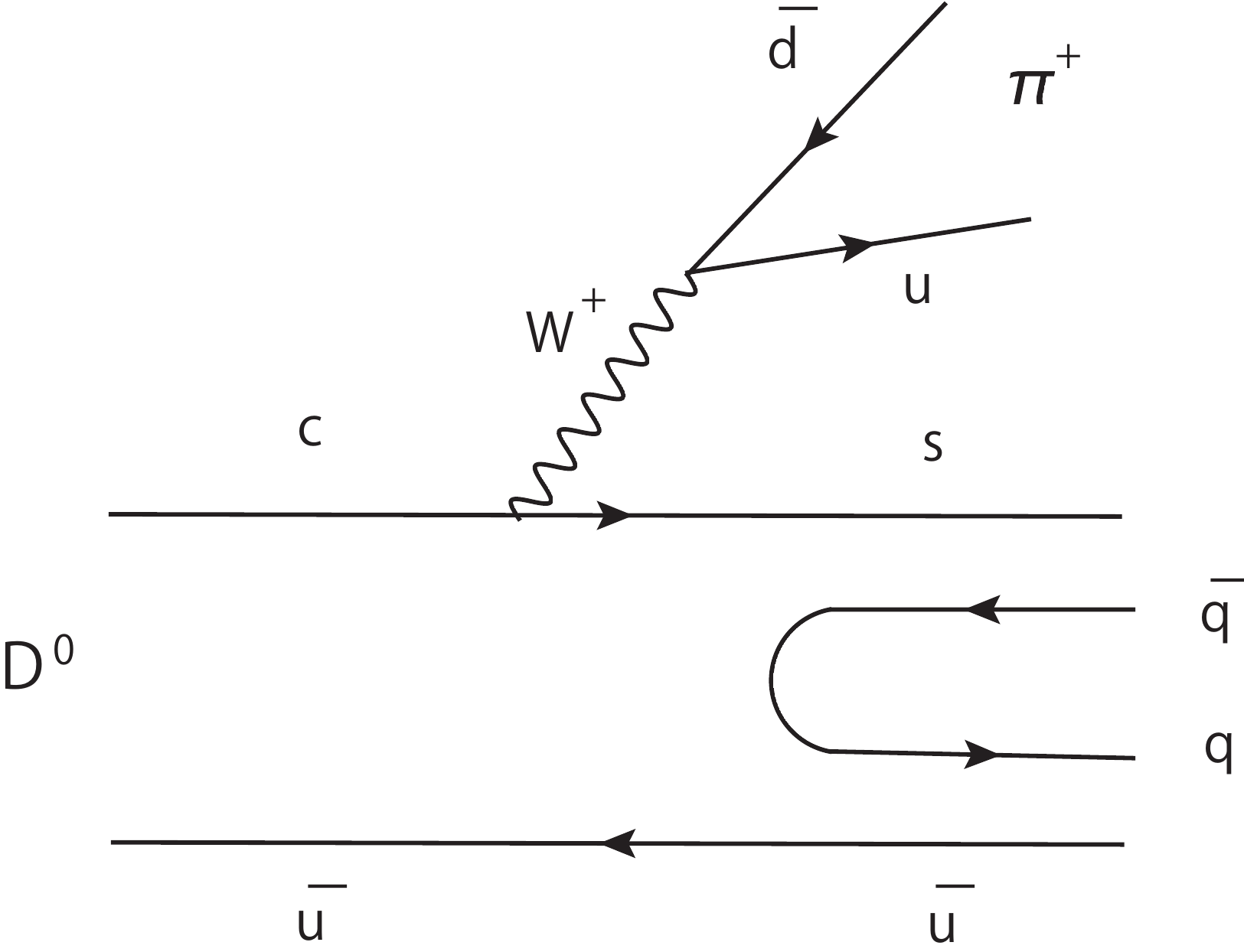}
\end{center}
\vspace{-0.7cm}
\caption{External emission of $D^0$ creating a $\pi^+$ and a $s\bar{u}$ pair, followed by hadronization of the $s\bar{u}$ pair. }
\label{Fig:1}
\end{figure}

As usual when studying a weak decay,  we start from the most favored Cabibbo mechanism at the quark level. For the $D^0 \to K^- \pi^+ \eta$ reaction we start with the external emission mechanism \cite{chau} shown in Fig. \ref{Fig:1}.

The $s\bar{u}$ pair can form a $K^-$ or a $K^{*-}$. The $K^{*-}$ could decay in $K^- \eta$ and then one would expect a signal of $\pi^+ K^{*-} \to \pi^+ K^-\eta$ with $K^- \eta$ in $p$-wave. A $K^*$ peak is actually visible in the experiment and provides the largest strength in the Dalitz plot \cite{belle}. However, the $K^-\eta$ threshold is about 150 MeV above the nominal $K^*$ mass, which has a width of 50 MeV. Hence, this contribution is negligible and we must attribute the experimental peak to a different reaction. In fact the $K^*$ peak is seen in the experiment in the $K\pi$ distribution. So, a different mechanism must be responsible for it, as we shall see below.\\  
 More difficult is to see how the $a_0(980)$ resonance, which also shows a large strength in the reaction \cite{belle}, can appear with this mechanism. The first step is to hadronize the $s\bar{u}$ component to form a pair of mesons. This is accomplished, as usual, introducing a $q\bar{q}$ pair with the quantum numbers of the vacuum. Here we are concerned about the flavor and then proceed as follow: A hadronic state $H$ is formed as
 
 \begin{equation}
 H= \sum_i s \bar{q}_i q_i \bar{u}= \left( M M \right)_{31}
 \end{equation}   
 where $M$ is the $q\bar{q}$ matrix. We then write the $M$ matrix in terms of the pseudoscalar mesons as
  
   \begin{equation}
  M \to P \equiv 
\left(  
\begin{array}{ccc}
\frac{\pi^0}{\sqrt{2}}+\frac{\eta}{\sqrt{3}}+\frac{\eta^\prime}{\sqrt{6}}
 & \pi^+ & K^+ \\
 
\pi^- & -\frac{\pi^0}{\sqrt{2}}+\frac{\eta}{\sqrt{3}}+\frac{\eta^\prime}{\sqrt{6}} & K^0  \\
K^- & \bar{K}^0 & -\frac{\eta}{\sqrt{3}}+\sqrt{\frac{2}{3}}\eta^\prime
\end{array}    
\right)\
\label{pmatrix}
 \end{equation} 
where the standard $\eta -\eta^\prime$ mixing has been assumed \cite{bramon}.  We find then

\begin{eqnarray}
H&=&K^- \left(\frac{\pi^0}{\sqrt{2}}+\frac{\eta}{\sqrt{3}}+\frac{\eta^\prime}{\sqrt{6}}
\right) + \bar{K}^0 \pi^- + \left(-\frac{\eta}{\sqrt{3}}+\sqrt{\frac{2}{3}}\eta^\prime\right) K^-\\
&\to& K^- \frac{\pi^0}{\sqrt{2}} + \bar{K}^0 \pi^-.
\end{eqnarray}

In the last step  above we see that the $K^-\eta$ state channel that we are looking for just cancels out. In addition we eliminate the $K^-\eta^\prime$ channel which is too far away for the relevant $K^*_0(700)$ resonance.\\
The $K^- \eta$ state has disappeared from tree level but we could obtain it through rescattering, $ \bar{K} \pi \to \bar{K}\eta$.
However, the $ \bar{K} \pi \to \bar{K}\eta$ through a $K^*$ resonance was found before to be an inefficient production method. We can try with $s$-wave. However, the $\bar{K}\eta$ threshold is around 1041 MeV, far away from the $K^*_0(700)$ peak, even considering the large $\kappa$ width. In addition, the coupling of the $\kappa$ to $K\eta$ is about half that of the $K\pi$ \cite{snd}.
All these things together indicate that the hadronization in this way, followed by rescattering to produce $K^-\eta$, is an inefficient way and this is corroborated by the experimental partial wave analysis which gives a very small contribution from $\bar{K}\eta$ in $s$-wave.

\begin{figure}[b!]
\begin{center}
\includegraphics[scale=0.4]{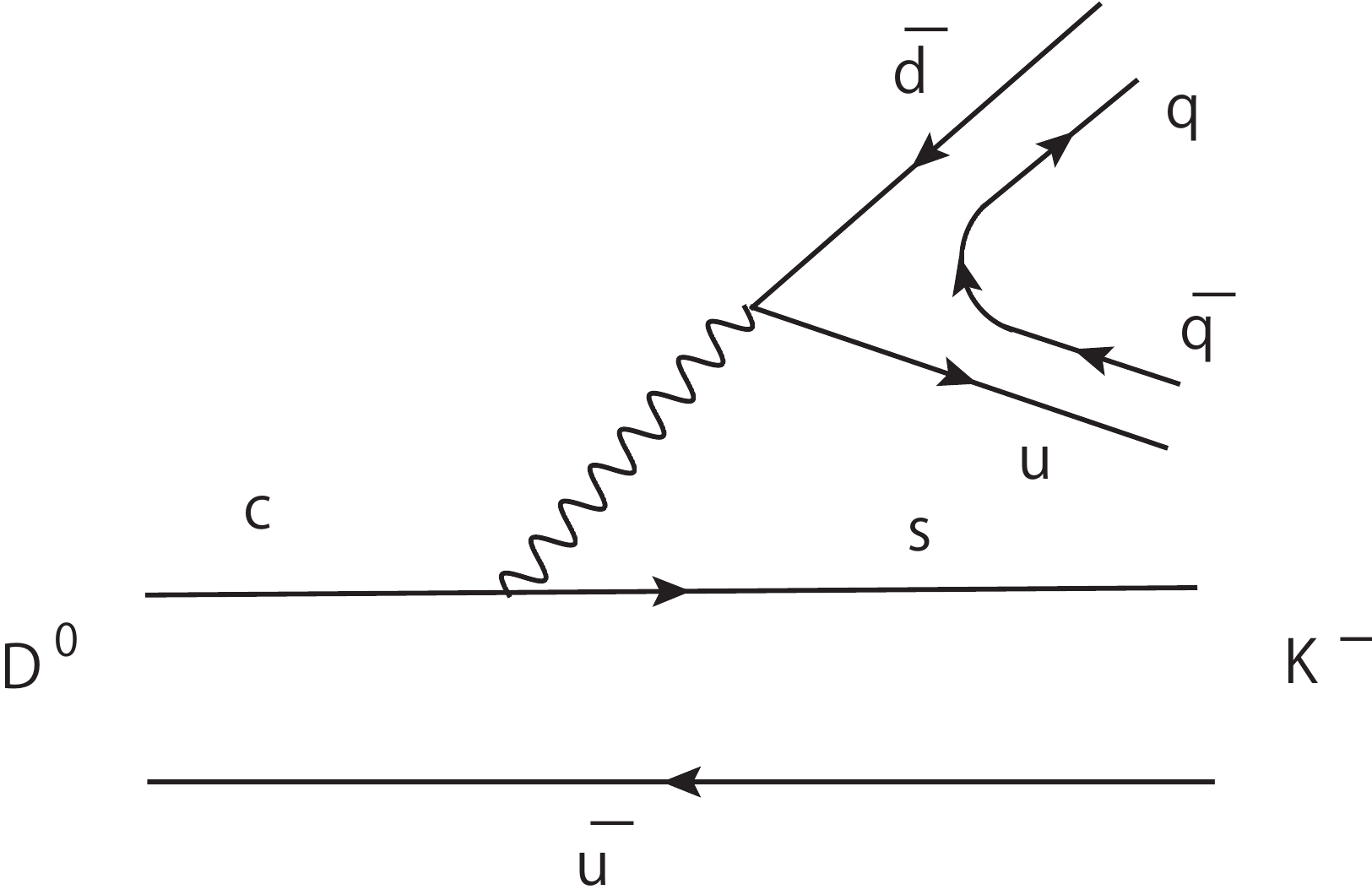}
\end{center}
\vspace{-0.7cm}
\caption{$D^0$ decay to $\bar{d} u K^-$, followed by hadronization of the $u\bar{d}$ pair.  }
\label{Fig:2}
\end{figure}

Next we resort to allowing the hadronization on the $d\bar{u}$ component as seen in Fig. \ref{Fig:2}, and use the $s\bar{u}$ component to produce the $K^-$. Following the same steps as before we find now:

\begin{eqnarray}
H&=& \sum_i u \bar{q}_i q_i \bar{d}= \left( P^2 \right)_{12}\nonumber\\
&=&\left(\frac{\pi^0}{\sqrt{2}}+\frac{\eta}{\sqrt{3}}+\frac{\eta^\prime}{\sqrt{6}}
\right) \pi^+ + \pi^+ \left(-\frac{\pi^0}{\sqrt{2}}+\frac{\eta}{\sqrt{3}}+\frac{\eta^\prime}{\sqrt{6}}\right) + K^+\bar{K}^0 \nonumber\\
&\to& \frac{2}{\sqrt{3}} \eta \pi^+ +  K^+\bar{K}^0
 \end{eqnarray}    

Here we see that the $\pi^0\pi^+$ channel has cancelled but not the $\eta\pi^+$, hence, together with the $K^-$ from the $s\bar{u}$ pair we have the hadronic final state
\begin{equation}
H^\prime = \frac{2}{\sqrt{3}} \eta \pi^+K^- +  K^+\bar{K}^0 K^-
\label{hprime}
 \end{equation}    
and we already have the $\eta \pi^+K^-$ final state.

\begin{figure}[b!]
\begin{center}
\includegraphics[scale=0.6]{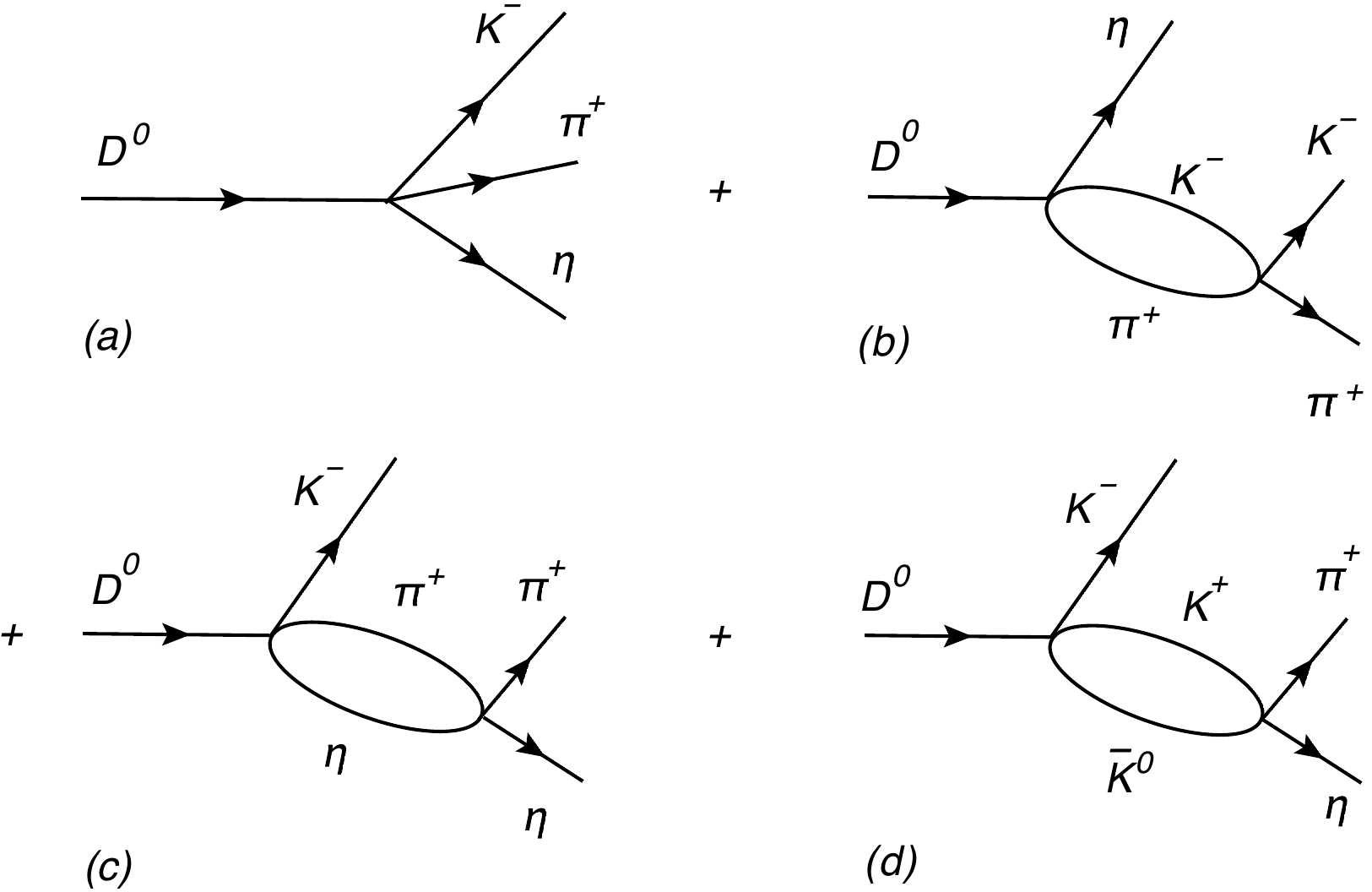}
\end{center}
\vspace{-0.7cm}
\caption{Final state interaction of the meson pairs.}
\label{Fig:3}
\end{figure}

The next step consists on taking into account the interaction of the meson pairs, which is depicted in Fig. \ref{Fig:3}.

We can have rescattering of the $K^-\pi^+$, which will produce the $\kappa$ and of the $K^+\bar{K}^0 \to \eta \pi^+$ which will produce the $a_0^+(980)$. For the reasons discussed above, we neglect the $\eta K^-$ scattering. Analytically we have:

\begin{eqnarray}
t &=C&  
\left\{
  h_{\eta \pi^+K^-}  + h_{\eta \pi^+K^-} 
\left[ 
 G_{K^-\pi^+}(M_{inv}({K^-\pi^+})) \ t_{K^-\pi^+,K^-\pi^+}(M_{inv}({K^-\pi^+}))
 \right.\right.\nonumber\\
&&\left.
 +  G_{\pi^+\eta}(M_{inv}({\pi^+\eta})) \ t_{\pi^+\eta,\pi^+\eta}(M_{inv}({\pi^+\eta}))
\right]\nonumber\\
&&
\left.
+h_{K^+ \bar{K}^0 K^-} G_{K^+\bar K}(M_{inv}({\pi^+\eta})) \ t_{K^+ \bar{K}^0,\pi^+\eta}(M_{inv}({\pi^+\eta}))
\right\}
\label{tinitial}
\end{eqnarray}

\noindent where $C$ is a global constant that will be taken from the normalization of the data and $h_i$ are the weights of the components in Eq. (\ref{hprime})
\begin{equation}
 h_{\eta \pi^+K^-}=\frac{2}{\sqrt{3}}\equiv h_1; \  \ h_{K^+ \bar{K}^0 K^-}=1\equiv h_3
 \label{weight1}
 \end{equation}    

The function $G_i$ and $t_i$ are the loop functions and scattering matrices respectively, which we take from \cite{npa,daixie} for the $\pi\eta$, $K\bar{K}$ channels and from \cite{dani,liang, guo} for the $K\pi$, $K\eta$ channels. As in \cite{daixie, liang}, the $G$ functions are regularized with a cut off, the maximum three momentum in the loop, with a value of $q_{max}\approx 600$ MeV.
Since in \cite{npa, daixie} one studies the neutral states, we mention here that, since in our isospin convention the $\pi^+$ is the $-|11>$ isospin state, then  

\begin{equation}
t_{K^+ \bar{K}^0,\pi^+\eta}=-t_{K \bar{K},\pi \eta}^{I=1}; \ \
t_{K^+ \bar{K}^0,\pi^+\eta}=\sqrt{2} t_{K^+ K^-,\pi^0\eta}
 \end{equation}    
As to the $K \pi$, $K \eta$ channels, we also take advantage to note that \cite{guo,liang} contain small correction terms with respect to \cite{bayar} and for completeness we give the detailed functions in the Appendix.

\begin{figure}[b!]
\begin{center}
\includegraphics[scale=0.8]{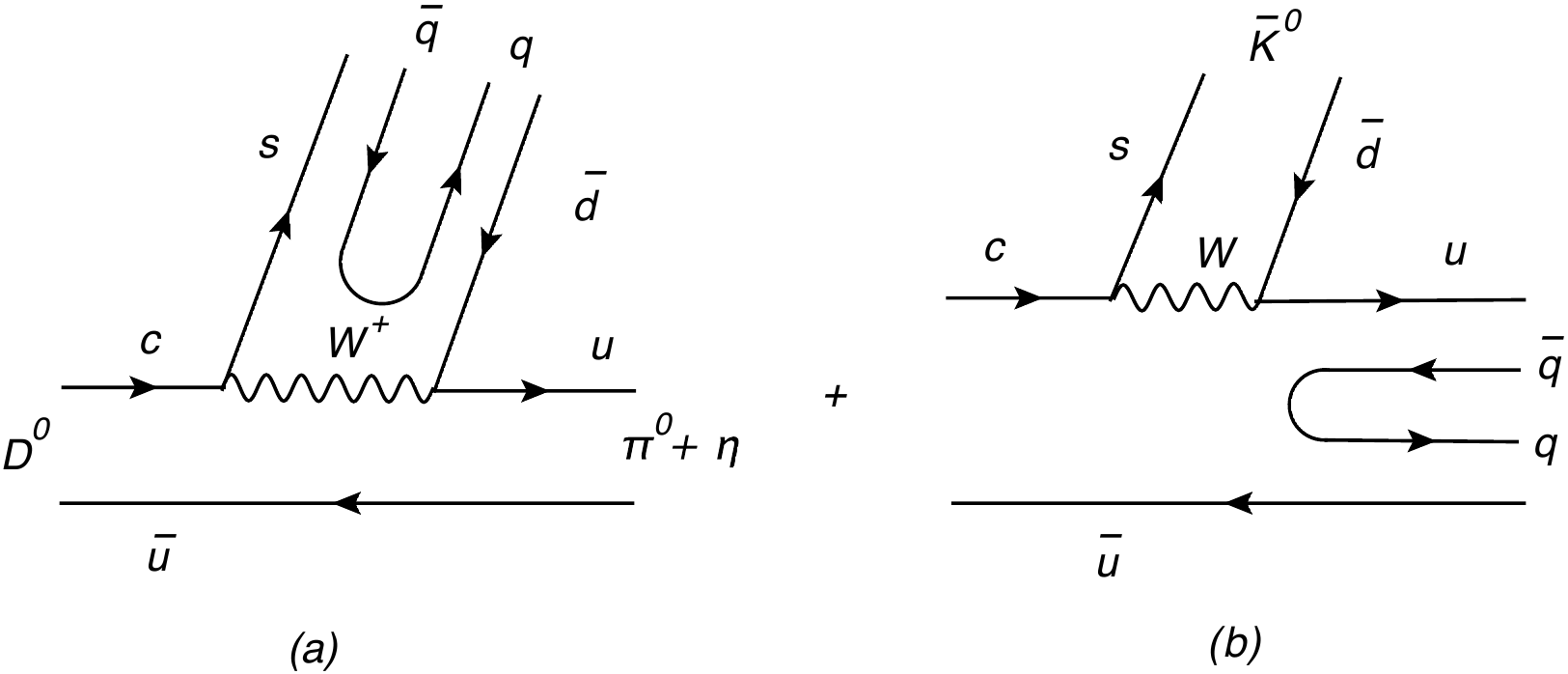}
\end{center}
\vspace{-0.6cm}
\caption{Internal emission for $D^0$ decay followed by hadronization.}
\label{Fig:4}
\end{figure}

So far we have relied on the most favored mechanism, color enhanced, external emission. There is also a possibility to reach the final state with internal emission, which is color suppressed, as depicted in Fig. \ref{Fig:4}.

We should first note that without hadronization we can produce $\bar{K}^{*0} \eta$ with $\bar{K}^{*0} \to \pi^+ K^-$ to which we shall come back. The $\pi^+K^-$ will be there in $p$-wave. For $s$-wave production we recur to hadronization. In the hadronization of the mechanism of Fig. \ref{Fig:4}(a) we will have the final state (omitting $\eta^\prime$)

\begin{eqnarray}
H&=& \sum_i s \bar{q}_i q_i \bar{d}= \left( P^2 \right)_{32}\nonumber\\
&=& K^- \pi^+ + \bar{K}^0 \left(-\frac{\pi^0}{\sqrt{2}}+\frac{\eta}{\sqrt{3}}\right)
-\frac{\eta}{\sqrt{3}} \bar{K}^0  \nonumber\\
&=&K^- \pi^+ -\bar{K}^0 \frac{\pi^0}{\sqrt{2}}
 \end{eqnarray}    

where the $\bar{K}^0 \eta$ channel has also cancelled. Including the $u\bar{u}$ state which is $\frac{\pi^0}{\sqrt{2}}+\frac{\eta}{\sqrt{3}}$, as seen in Eq. (\ref{pmatrix}), we have

\begin{equation}
H^\prime =\left(K^-\pi^+ -\bar{K}^0 \frac{\pi^0}{\sqrt{2}}\right)\left( \frac{\pi^0}{\sqrt{2}}+\frac{\eta}{\sqrt{3}}\right)
\label{hprime2}
 \end{equation} 
The mechanism of Fig. \ref{Fig:4}(b) leads to the hadronized state

\begin{eqnarray}
H^\prime&=&  \bar{K}^0 \sum_i u \bar{q}_i q_i \bar{u}= \bar{K}^0  \left( P^2 \right)_{11}\nonumber\\
&=& \bar{K}^0
\left( \left(\frac{\pi^0}{\sqrt{2}}+\frac{\eta}{\sqrt{3}}\right)^2 + \pi^+\pi^- + K^+ K^-
\right)
 \end{eqnarray}    

We can see that several channels are produced with both mechanisms and we add the two contributions

\begin{eqnarray}
H^\prime &=& 
K^-\pi^+ \frac{\pi^0}{\sqrt{2}} +K^- \pi^+ \frac{\eta}{\sqrt{3}}
+\bar{K}^0 \frac{\pi^0 \eta}{\sqrt{6}} +\bar{K}^0 \frac{\eta\eta}{3 }\nonumber\\
&&+\bar{K}^0\pi^+\pi^- +\bar{K}^0 K^+ K^-
 \end{eqnarray}  

The $\bar{K}^0\pi^0\pi^0$ combination disappears in the sum. We see that we have a tree level contribution in $K^- \pi^+ \eta$, and several other terms from where we can obtain the final state $K^- \pi^+ \eta$ with rescattering. But some of these channels are useless to produce the final state. For instance, the $K^- \pi^+\pi^0$ term. The $\pi^+\pi^0$ in $s$-wave can be in $I=2$ ($I=0$ is not allowed because $I_3 = 1$), but not in $I=1$, and hence cannot create the $\pi^+\eta$. The $K^-\pi^0 \to K^- \eta$ only sees the tail of the $\kappa$, as we have discussed previously, and hence, we disregard this channel. The $\bar{K}^0\eta\eta$ is equally unsuited since $K^0\eta \to K^- \pi^+$ will also only see the tail of the $\kappa$. For the same reason $\bar{K}^0 \pi^+\pi^-$ is also unsuited since $K^0\pi^- \to K^- \eta$
will also only see the $\kappa$ resonance tail. Hence for practical purposes we are left to a hadronic state

\begin{equation}
H^\prime_{int} = 
\bar{h}_1 K^-\pi^+ \eta
+ \bar{h}_2 \bar{K}^0 \pi^0 \eta
+ \bar{h}_3 \bar{K}^0 K^+ K^-
 \end{equation}    
with
\begin{equation}
\bar{h}_1=\frac{1}{\sqrt{3}}; \  \ \bar{h}_2=\frac{1}{\sqrt{6}};  \ \ \bar{h}_3=1
\label{weight2}
 \end{equation}    
 
\begin{figure}[b!]
\begin{center}
\includegraphics[scale=0.5]{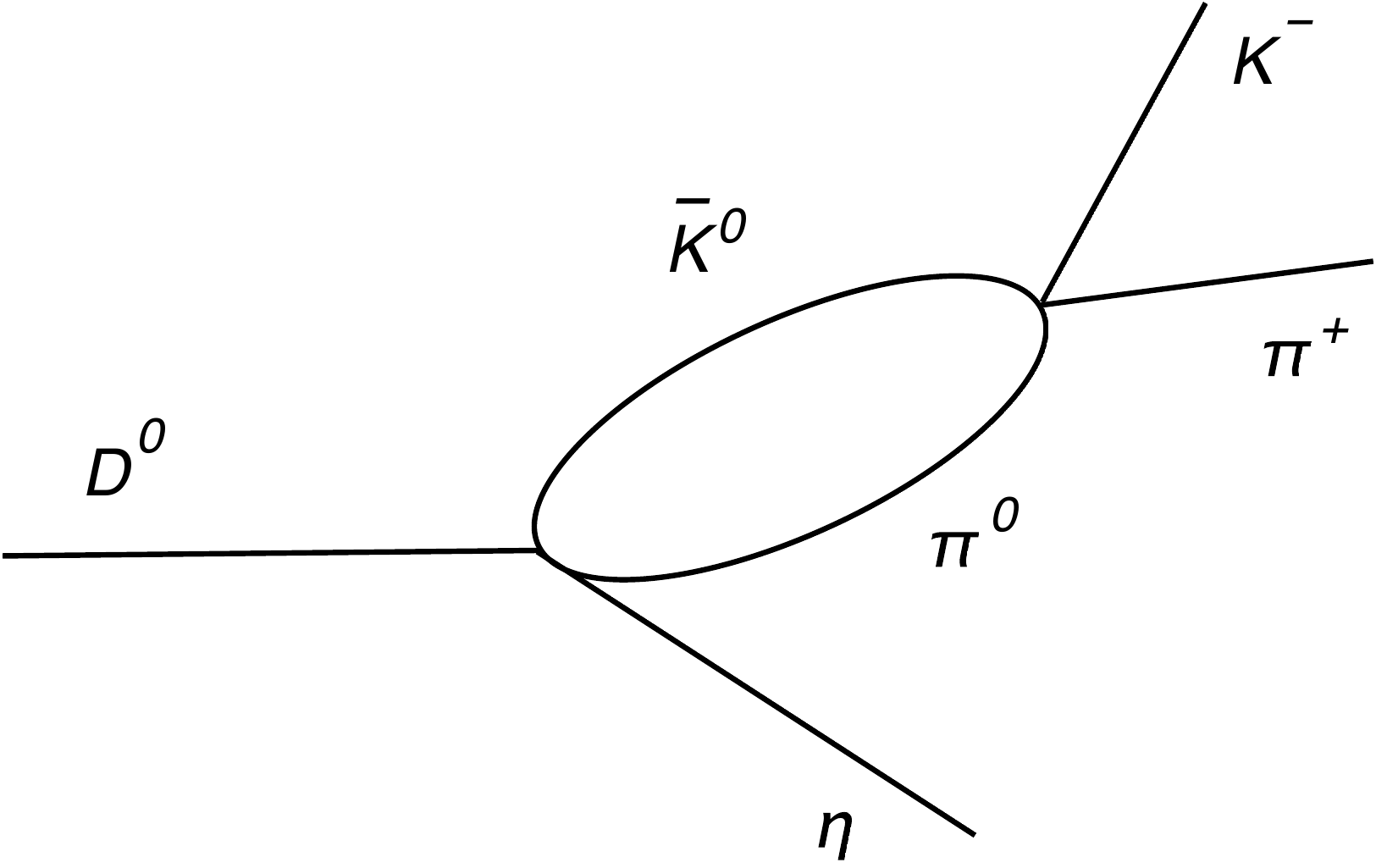}
\end{center}
\vspace{-0.7cm}
\caption{Final state interaction of the $\bar{K}^0 \pi^0$ pair in the $\bar{K}^0 \pi^0 \eta$ term.}
\label{Fig:5}
\end{figure}

We see that the states $K^- \pi^+ \eta$ and $\bar{K}^0 K^+ K^-$ also appeared in external emission Eq. (\ref{hprime}). There is a new term $\bar{h}_2 \bar{K}^0 \pi^0 \eta$ and we can have $\bar{K}^0 \pi^0 \to K^- \pi^+ $ reaching the final $K^- \pi^+\eta$ state, see Fig. \ref{Fig:5}. The internal emission term should have a different weight, $C\beta$, with the modulus of $\beta$ smaller than 1. We will use this small term for only fine tuning. 

The amplitude for the $K^-\pi^+ \eta$ production process including rescattering of the different terms is then given by ( see Eqs. (\ref{weight1}) and (\ref{weight2}) for the $h_i$ and $\bar{h}_i$ coefficients)

\begin{eqnarray}
t &=C&  
\left\{
  h_1  + \beta \bar{h}_1
+ G_{K\pi}(M_{inv}({K^-\pi^+})) 
\left[ (h_1  + \beta \bar{h}_1) \ t_{K^-\pi^+,K^-\pi^+}(M_{inv}({K^-\pi^+}))
 \right.
 \right.\nonumber\\
&&
\left.+ \beta \bar{h}_2 \ t_{\bar{K}^0\pi^0,K^-\pi^+}(M_{inv}({K^-\pi^+}))
\right]
 +  (h_1  + \beta \bar{h}_1) \ G_{\pi\eta}(M_{inv}({\pi^+\eta})) \ t_{\pi^+\eta,\pi^+\eta}(M_{inv}({\pi^+\eta}))
\nonumber\\
&&
\left.
+(h_3  + \beta \bar{h}_3) \ G_{K\bar{K}}(M_{inv}({\pi^+\eta})) \ t_{K^+ \bar{K}^0,\pi^+\eta}(M_{inv}({\pi^+\eta}))
\right\}
\label{tfinal}
\end{eqnarray}

As in \cite{vinicius} (see Eq.~(19) of Ref.~\cite{vinicius}) we smoothly extrapolate the $Gt$ amplitude above an energy $M_{cut}=1100$ MeV, and the  results barely change for different sensible extrapolations.

\subsection{The $D^0 \to \eta  \bar{K}^{*0} \to \pi^+ K^- \eta$ contribution }

\begin{figure}[b!]
\begin{center}
\includegraphics[scale=0.5]{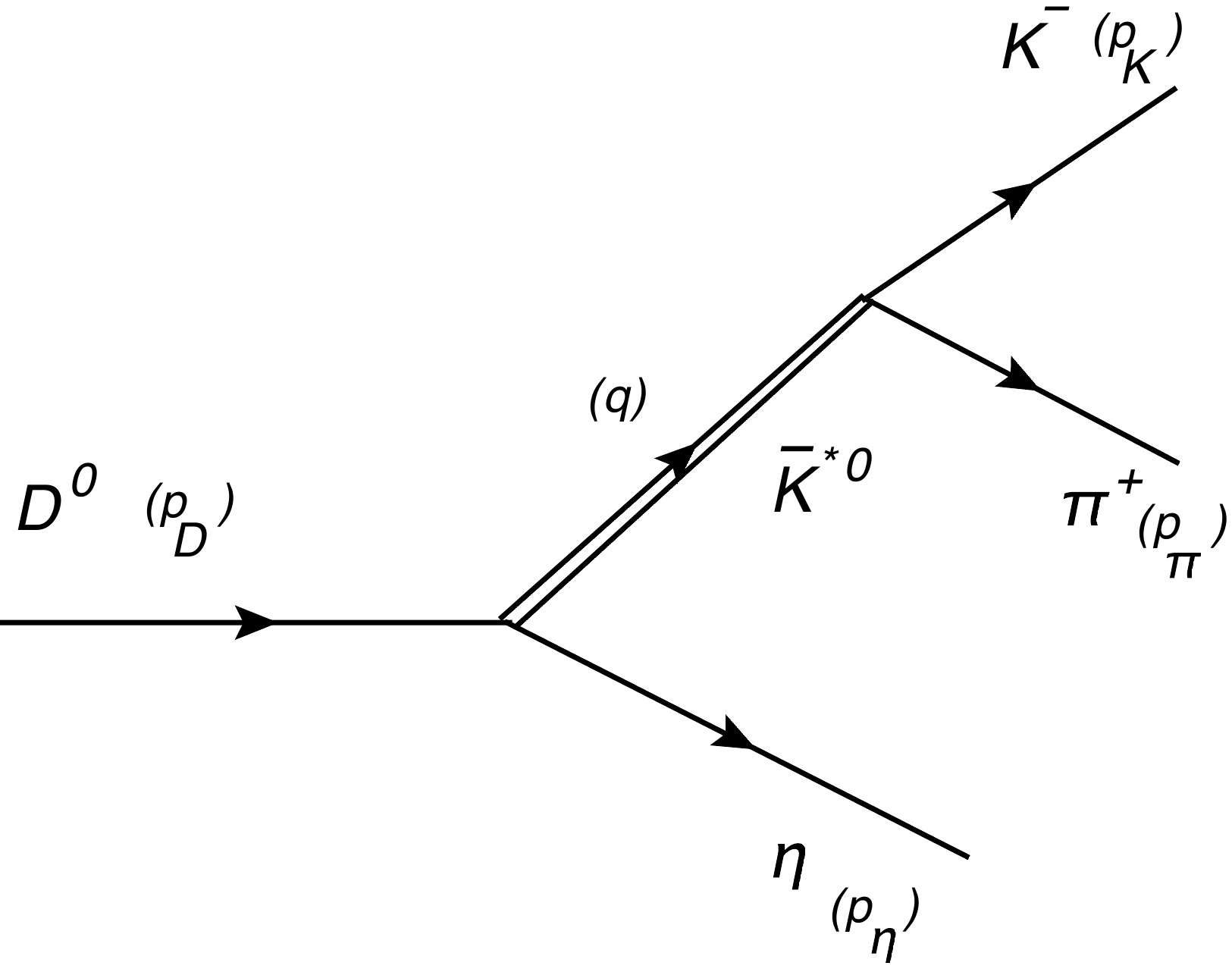}
\end{center}
\vspace{-0.7cm}
\caption{Diagram for $D^0 \to  \bar{K}^{*0} \eta \to K^- \pi^+  \eta$. The momenta of the particles are written in parenthesis and $q\equiv p_K + p_\pi$}
\label{Fig:6}
\end{figure}

We saw in connection with Fig. \ref{Fig:1} that we could produce $\pi^+ K^{*-}$ with external emission. Then the $K^{*-}$ could decay to $K^-\eta$ in $p$-wave, but the process was inefficient since it involved the tail of the $K^*$ far away from the nominal $K^*$ mass. However, the mechanisms of internal emission in Fig. \ref{Fig:4} can both produce ${\bar K}^{*0} \eta$, and the ${\bar K}^{*0}$ can decay to $K^- \pi^+$. We derive here the amplitude for the $D^0 \to \eta  \bar{K}^{*0} \to \pi^+ K^- \eta$ process, which  will add incoherently to the $s$-wave contributions that we have studied before. The mechanism of production is depicted in Fig. \ref{Fig:6}.

Up to an unknown constant $D$, which we will fit to the experimental strength, the full relativistic amplitude, needed to see the contribution of the mechanism in a large invariant mass span, is given by

\begin{eqnarray}
\cal{M}&=&D(p_D+ p_\eta)^\mu 
\left(-g_{\mu\nu}+ \frac{q_\mu q_\nu}{M_{K^*}^2}\right)
(p_K-p_\pi)^\nu \frac{1}{q^2- M_{K^*}^2 + i M_{K^*} \Gamma_{K^*}}\nonumber\\
&=& \frac{D}{q^2- M_{K^*}^2 + i M_{K^*} \Gamma_{K^*}}
\left[
-(p_D+ p_\eta)\cdot (p_K-p_\pi)+  \frac{(p_K+ p_\pi)\cdot (p_K-p_\pi)}{M_{K^*}^2}  (p_D+ p_\eta)\cdot (p_K + p_\pi)
\right]\nonumber
 \end{eqnarray}    
Using
$(p_K+ p_\pi)\cdot (p_K-p_\pi)=m_K^2-m_\pi^2$ and labelling the particles $K^-(1)$, $\pi^+(2)$, $\eta(3)$, we write $s_{13}=(p_K+ p_\eta)^2$; $s_{23}=(p_\pi+ p_\eta)^2$ and the transition matrix $\cal{M}$ can be written as

\begin{equation}
{\cal M} =\frac{D} {q^2- M_{K^*}^2 + i M_{K^*} \Gamma_{K^*}}
\left[
(m_K^2-m_\pi^2)  \frac{(m_D^2-m_\eta^2)}{M_{K^*}^2} - s_{13}+ s_{23} 
\right]
\label{MKstar}
 \end{equation}    

Since the $s$-wave terms in $t$ in Eq. (\ref{tfinal}) and $\cal{M}$ do not interfere in the angle integrated distributions we define

\begin{equation}
|t^\prime|^2= |t|^2+ |{\cal M}|^2
\end{equation}    

\noindent and then $|t^\prime|^2$ depends on $s_{12}=M_{inv}^2(K^-\pi^+)$,  $s_{13}=M_{inv}^2(K^-\eta)$, $s_{23}=M_{inv}^2(\pi^+\eta)$, although only two of these variables are independent since 
\begin{equation}
s_{12}+s_{13}+s_{23}=m_D^2+m_K^2+m_\pi^2+m_\eta^2
\label{invariants}
\end{equation}   
Then we use the formula of the PDG for three body decay \cite{pdg}
\begin{equation}
\frac{d^2 \Gamma}{dM_{inv}^2(12) dM_{inv}^2(23)}=\frac{1}{(2\pi)^3}
\frac{1}{32 m_D^3} |t^\prime|^2
\end{equation}    

\noindent and we integrate over either of the invariant masses to obtain the single invariant mass distributions. Permuting the indices 123 and using Eq. (\ref{invariants}) we easily find $d\Gamma/dM^{2}_{inv}(13)$.

\section{Results}

\begin{figure}[b!]
\begin{center}
\includegraphics[scale=0.8]{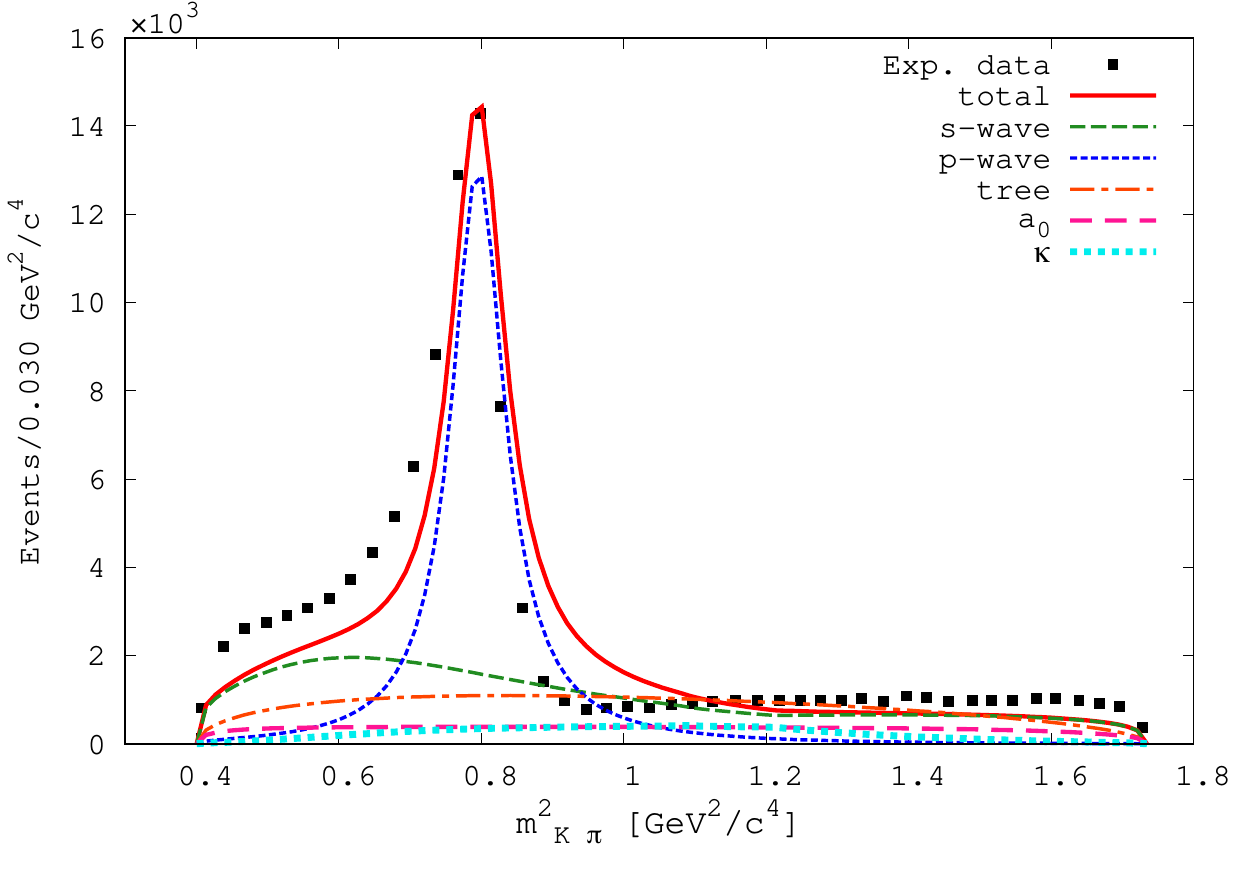}
\end{center}
\vspace{-0.7cm}
\caption{$M_{K\pi}$ distribution. Individual contributions. The total $s$-wave contains the tree level, the $a_0(980)$ and the $\kappa$ rescattering terms.}
\label{Fig:7}
\end{figure}

\begin{figure}[b!]
\begin{center}
\includegraphics[scale=0.8]{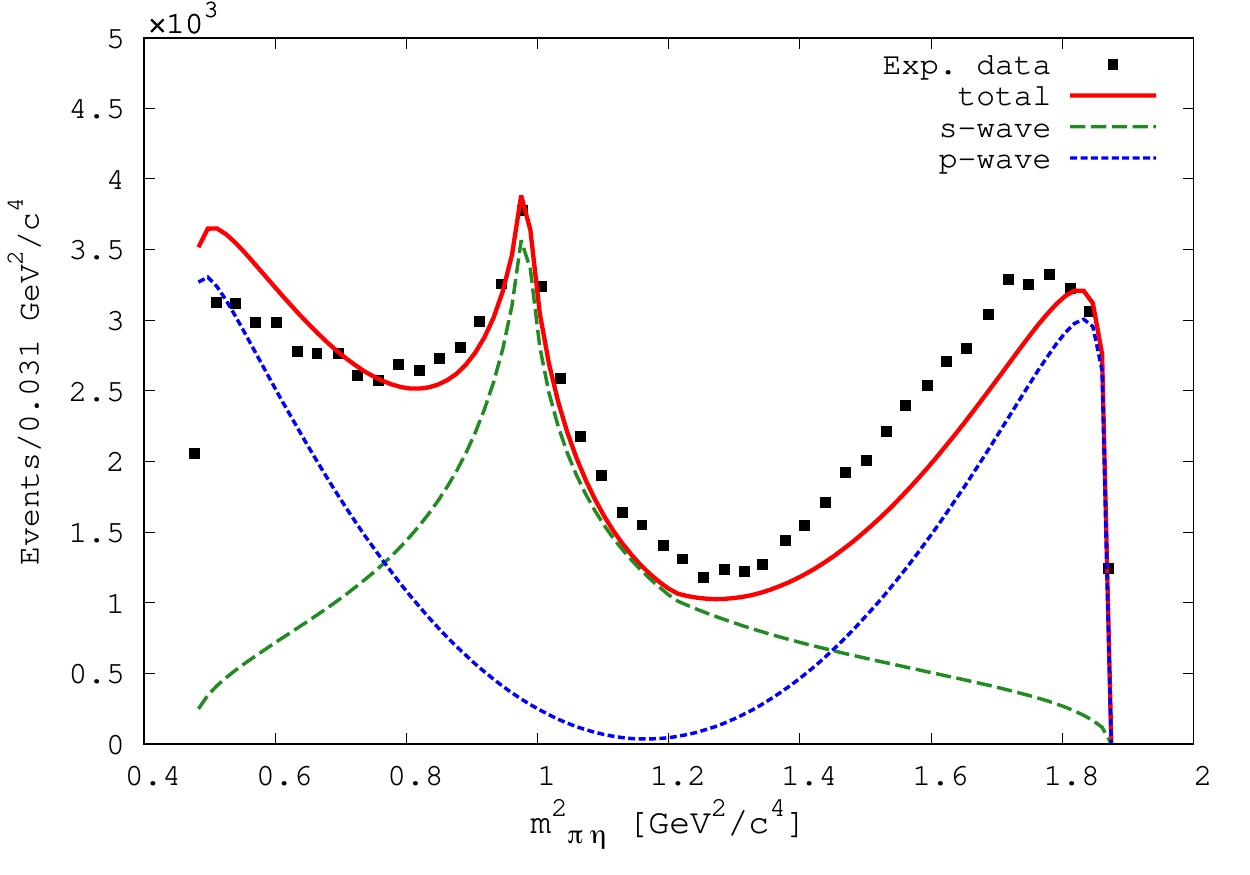}
\end{center}
\vspace{-0.7cm}
\caption{$M_{\pi\eta}$ distribution.}
\label{Fig:8}
\end{figure}

\begin{figure}[b!]
\begin{center}
\includegraphics[scale=0.8]{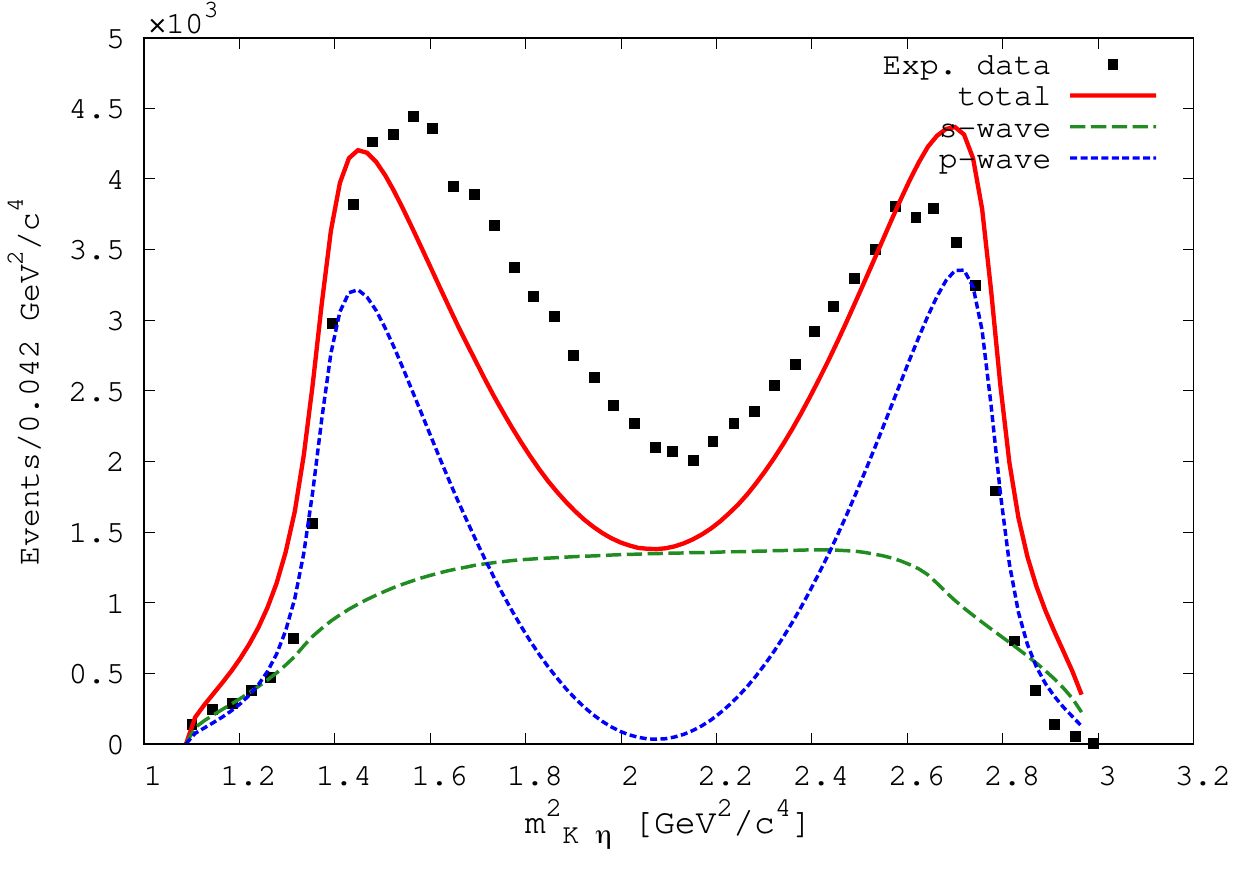}
\end{center}
\vspace{-0.7cm}
\caption{$M_{K\eta}$ distribution.}
\label{Fig:9}
\end{figure}

We have two parameters at our disposal to fit the data if we consider only the dominant, external emission mechanism, $C$ and $D$. They are uncorrelated since $C$ determines the absolute strength of the width and $D$ controls the strength of the $K^*$ excitation. In Figs. \ref{Fig:7}, \ref{Fig:8}, \ref{Fig:9} we show the results obtained for the three invariant mass distributions using only the $C$ and $D$ parameters. The corresponding values are $C=1.0$ and $D=0.05$.

\begin{figure}[b!]
\begin{center}
\includegraphics[scale=0.8]{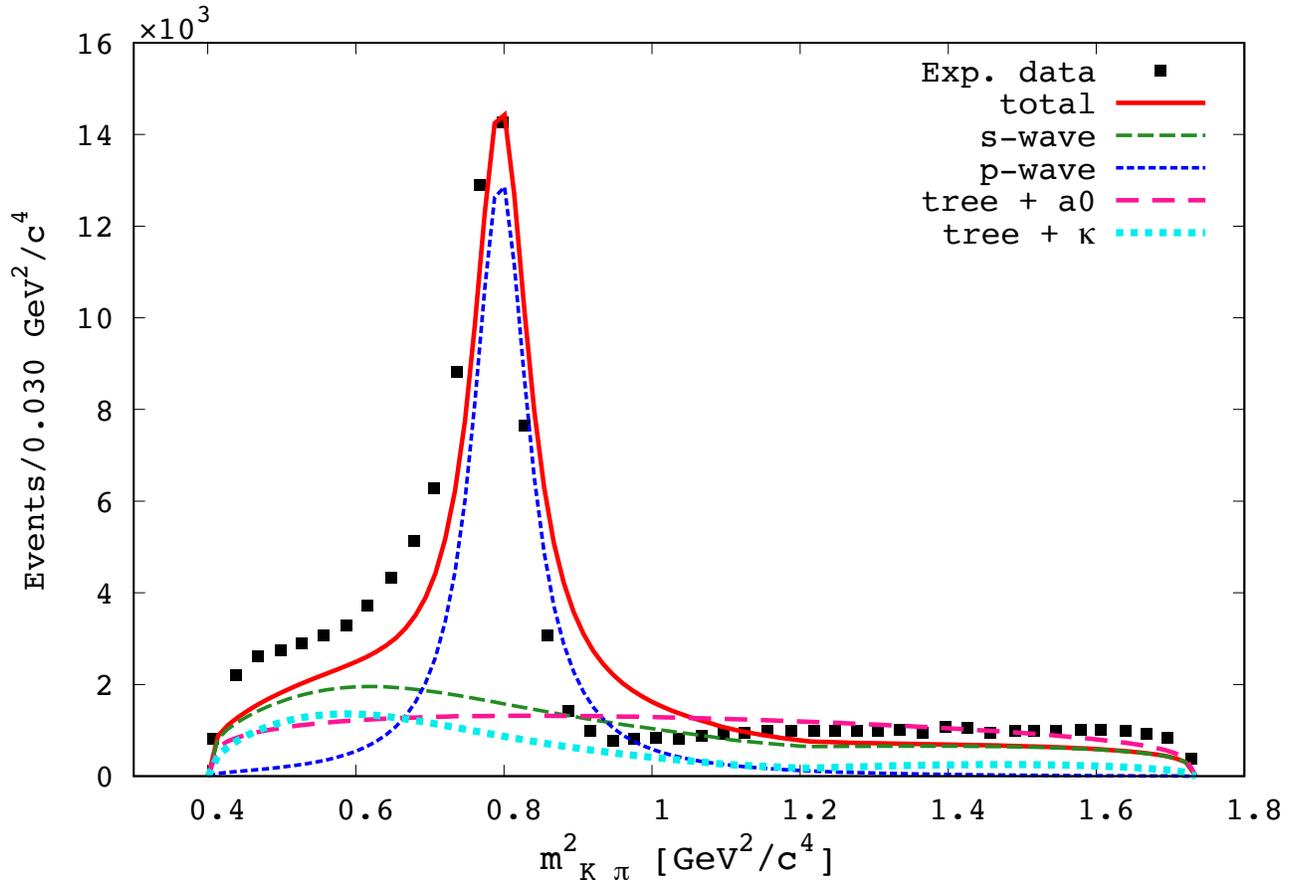}
\end{center}
\vspace{-0.7cm}
\caption{$M_{K\pi}$ distribution. Combined contributions. The total $s$-wave contains the tree level plus the $a_0(980)$ and the $\kappa$ rescattering terms.}
\label{Fig:7.new}
\end{figure}

We can see that we get a good reproduction of the data at a qualitative level. We reproduce, because it is an input, the peak of the $m_{K\pi}$ mass distribution. What is a consequence  of our theoretical formalism is the accumulated strength below the peak of the $\bar{K}^{*0}$ resonance. 
To see that, we show in Fig.~\ref{Fig:7} the contributions of the $a_0(980)$ (the two terms of Eq.~(\ref{tinitial}) involving the $ t_{\pi^+\eta,\pi^+\eta} $ and $t_{K^+ \bar{K}^0,\pi^+\eta} $ amplitudes) and $\kappa$ (term of Eq.~(\ref{tinitial}) involving $t_{K^-\pi^+,K^-\pi^+}$). We should note that the $t_{K^-\pi^+,K^-\pi^+}$ amplitude contains contributions from $I=1/2$ (the $\kappa$) and $I=3/2$, but the $I=1/2$ is dominant and we shall call this the $\kappa$ contribution. As we see, both the $a_0(980)$ and $\kappa$ contributions are small compared to the contribution of the tree level (first term of Eq.~(\ref{tinitial})). However, upon interference with the tree level, the effect of the $a_0(980)$ and $\kappa$ get reinforced. This is better seen in Fig.~\ref{Fig:7.new}, where we show separately the contributions, tree$+a_0(980)$, tree$+\kappa$, and tree$+a_0(980)+\kappa$ ($s$-wave). What the two figures tell us is the importance of the tree level term, enhancing the contributions of the $a_0(980)$ and $\kappa$ through interference. It is thus clear that a proper analysis of the data will require the explicit consideration of the tree level in order to extract the $s$-wave $\pi K$ and $\pi \eta$ amplitude from them.  
 It is also striking that the prominent role of the $a_0(980)$ in the $\pi\eta$ mass distribution is obtained in our approach without introducing it in the formalism, unlike the $\bar{K}^{*0}$ contribution which is put by hand. This comes as a consequence of the rescattering of $\pi^+\eta$ and $K^+\bar{K}^{0}$, as seen in Fig. \ref{Fig:3}. The scattering amplitude  $\pi^+\eta \to \pi^+\eta$ and  $K^+\bar{K}^{0} \to \pi^+\eta$ in the chiral unitary approach contain the $a_0(980)$ resonance, which comes as a consequence of the interaction of the mesons and is also not introduced by hand in the approach. We should stress the cusp like shape of this resonance both in the theory and in the experiment, something already noted in the high statistics BESIII experiment on the $\chi_{c1} \to  \eta \pi^+\pi^-$ reaction \cite{beschi1} accurately described theoretically in \cite{liangxie} along similar lines as shown here. It is also interesting to see the curious effect that the $\bar{K}^{*0}$ contribution has in the $M_{\pi\eta}$ and $M_{K\eta}$ distributions, giving rise to two broad peaks at lower and higher invariant masses. We must note that this is also reproduced in our approach but it requieres the use of the full relativistic amplitude of Eq. (\ref{MKstar}) and is easily missed in nonrelativistic approximations. These peaks, correctly interpreted in the experimental analysis of \cite{besiii} to the light of our different formulation, are typical examples of replicas of invariant mass distributions of resonant peaks of one particular invariant mass. It is important to identify them correctly to avoid claims of new resonances. We can see in these plots that the effect of the $a_0(980)$ and $\kappa$ resonances are instead rather smooth and structureless in the non resonant invariant plots.\\
Finally, since the $C$ coefficient governs the absolute normalization and the $D$ coefficient the strength of the $\bar{K}^{*0}$, the relative strength between the $a_0(980)$ peak and the low energy $\bar{K}\pi$ bump is a prediction of the theory with no free parameters.

\subsection{Effect of the internal emission mechanism}

\begin{figure}[b!]
\begin{center}
\includegraphics[scale=0.8]{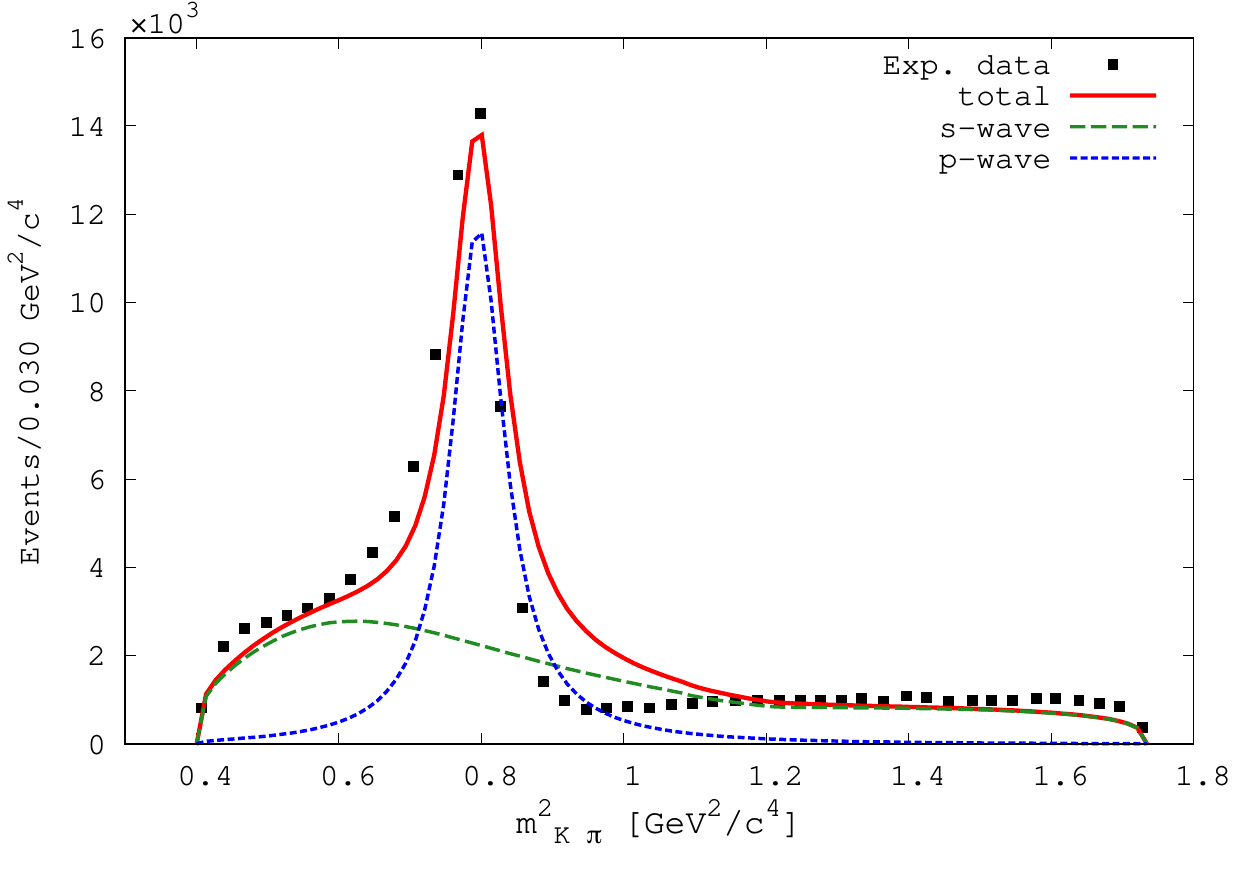}
\end{center}
\vspace{-0.7cm}
\caption{$M_{K\pi}$ distribution including internal emission.}
\label{Fig:10}
\end{figure}

\begin{figure}[b!]
\begin{center}
\includegraphics[scale=0.8]{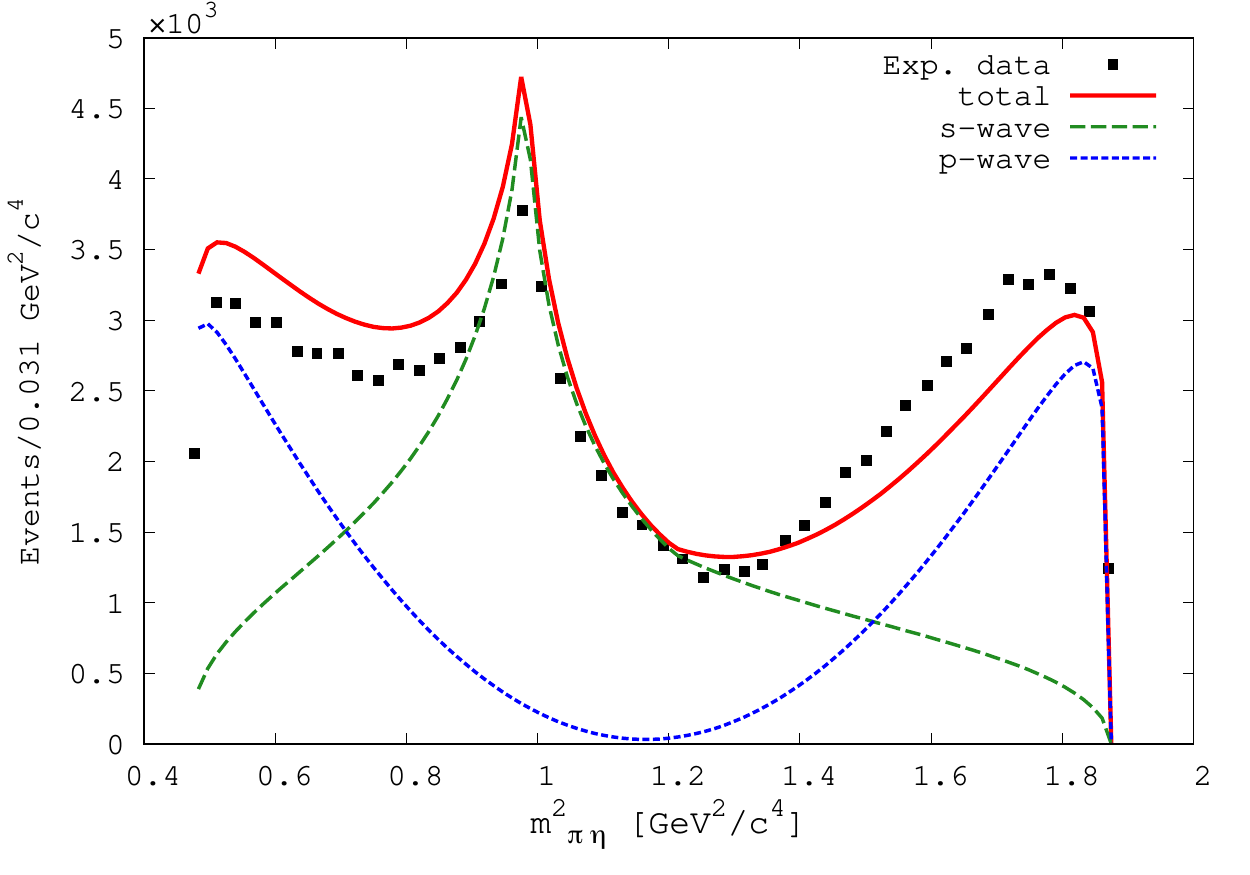}
\end{center}
\vspace{-0.7cm}
\caption{$M_{\pi\eta}$ distribution including internal emission.}
\label{Fig:11}
\end{figure}

\begin{figure}[b!]
\begin{center}
\includegraphics[scale=0.8]{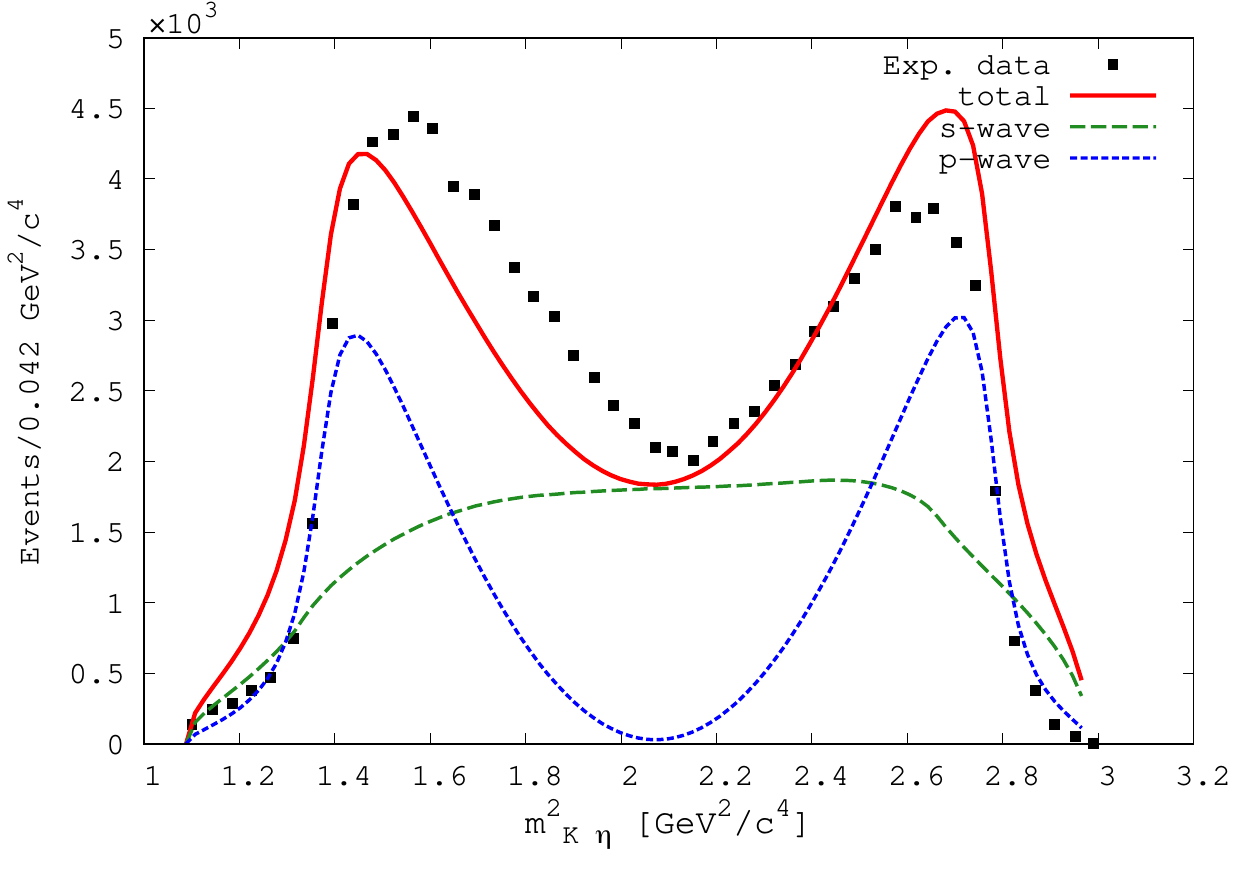}
\end{center}
\vspace{-0.7cm}
\caption{$M_{K\eta}$ distribution including internal emission.}
\label{Fig:12}
\end{figure}

The agreement with data obtained in Figs. \ref{Fig:7}, \ref{Fig:8}, \ref{Fig:9} is fair considering that only the global strength and that of the $\bar{K}^*_0$ peak have been fitted to the data. It is also unnecessary to demand a better agreement with the data that are not efficiency corrected  \cite{longke}, something common in Belle data (see comments in \cite{wangliang}). Yet, in a similar range of energies, efficiency corrections tend to be similar and, with this perspective and the due caution, we try to improve the agreement with data in the low energy $\bar{K}\pi$ mass distribution and the $a_0(980)$ peak, using the contribution from the internal emission mechanism (Fig. \ref{Fig:4}). We have at our disposal just one new parameter, $\beta$, which should be small compared to unity as we have already discussed. Since the internal emission is suppressed by a color factor with respect to external emission, we should expect the modulus of $\beta$ to  be of the order of 1/3.
In Figs. \ref{Fig:10}, \ref{Fig:11}, \ref{Fig:12} we show  the results with a fit with the values of the parameters $C=1.5$, $D=0.03$, $\beta=-0.4$.
The agreement with the data improves a bit, particularly the simultaneous reproduction of the strength of the $\bar{K}\pi$ distribution at low energies and of the $a_0(980)$ peak, which are the genuine predictions of the theory, but the general trend was already reproduced by the dominant external emission mechanism, where for the $s$-wave we only had the global strength as a degree of freedom.

\section{Conclusions}
  We have studied the $D^0 \to K^- \pi^+ \eta$ decay, recently measured by the Belle collaboration, and found it to be very well suited to provide valuable information on the scalar mesons $a_0(980)$ and $\kappa$ ($K^*_0(700)$). The analysis is done studying first how the primary quark production proceeds and then hadronizing pairs of quarks to provide two pseudoscalar mesons. We find that while the $K^- \pi^+ \eta$ state can be produced in a primary stage, prior to any final state interaction consideration, the interaction of mesons, and not only the final ones, gives rise to two resonances, the $\kappa$ in the final $\bar{K} \pi$ channel and the $a_0(980)$ in the final $\pi^+ \eta$ channel. Our formalism, which uses the chiral unitary approach to account for the interaction of pairs of pseudoscalar mesons, is well suited for these kind of reactions. It produces simultaneously the two resonances and provides their relative strength with no free parameters in the dominant mode of decay, based on external emission. A small fraction of internal emission is also taken into account in the approach, leading to a better agreement with the data. Including empirically the $\bar{K}^{*0} \eta$ production we find a relatively good agreement with the data in the three invariant mass distributions and all the range of masses.  The agreement found with the data gives support to our theoretical scheme, where the final state interaction is responsible for the main features, and indirectly to the nature of the resonances $\kappa$ and $a_0(980)$, which do not qualify as $q \bar{q}$ states, but come as a consequence of the interaction of the mesons pairs in coupled channels. Together with the success obtained in other reactions using the same idea, the information favoring this picture is piling up, revealing the different nature of the low lying scalar mesons from the ordinary $q \bar{q}$ mesons.

\section{ACKNOWLEDGEMENT}
G. T. acknowledges the support of PASPA-DGAPA, UNAM for a sabbatical leave. 
The work of N. I. was partly supported by JSPS Overseas Research Fellowships and JSPS KAKENHI Grant Number JP19K14709.
This work is partly supported by the Spanish Ministerio de Economia y Competitividad and European FEDER funds under Contracts No. FIS2017-84038-C2-1-P B and No. FIS2017-84038-C2-2-P B. This project has received funding from the European Unions Horizon 2020 research and innovation programe under grant agreement No 824093 for the **STRONG-2020 project.

\appendix*
\section{Scattering amplitude in the $K\pi$, $K\eta$ channels}
The $T$ matrix is taken in matrix form as
\begin{equation}
T= [ 1- V G]^{-1} V
\end{equation}
with the $\pi^- K^+(1)$, $\pi^0 K^0(2)$, $\eta K^0 (3)$ channels and we have

\begin{eqnarray}
V_{11} &=&\frac{-1}{6f^2} \left( \frac{3}{2}s- \frac{3}{2s}(m_\pi^2-m_K^2)^2 \right)\\
V_{12}&=&\frac{1}{2 \sqrt{2}f^2} \left( \frac{3}{2}s- m_\pi^2-m_K^2-  \frac{(m_\pi^2-m_K^2)^2}{2s} \right)\\
V_{22} &=&\frac{-1}{4f^2} \left( -\frac{s}{2}+ m_\pi^2+m_K^2-  \frac{(m_\pi^2-m_K^2)^2}{2s} \right)\\
V_{13}&=&\frac{1}{2\sqrt{6}f^2} \left( \frac{3}{2}s- \frac{7}{6}m_\pi^2-\frac{1}{2}m_\eta^2-\frac{1}{3}m_K^2+ \frac{3}{2s}(m_\pi^2-m_K^2)(m_\eta^2-m_K^2) \right)\\
V_{23}&=& -\frac{1}{4\sqrt{3}f^2} \left( \frac{3}{2}s- \frac{7}{6}m_\pi^2-\frac{1}{2}m_\eta^2-\frac{1}{3}m_K^2+ \frac{3}{2s}(m_\pi^2-m_K^2)(m_\eta^2-m_K^2) \right)\\
V_{33}&=& -\frac{1}{4f^2} \left(- \frac{3}{2}s- \frac{2}{3}m_\pi^2+m_\eta^2+ 3m_K^2- \frac{3}{2s}(m_\eta^2-m_K^2)^2 \right)
\end{eqnarray}
 
where $f$ is the pion decay constant, $f=93$ MeV, and $s$ the square of the center of mass energy.

\end{document}